# Intrusion Detection Systems Using Support Vector Machines on the KDDCUP’99 and NSL-KDD Datasets. A Comprehensive Survey

Mikel K. Ngueajio[1], Gloria Washington[1], Danda B. Rawat[1], and Yolande Ngueabou[2]

[1] Howard University, Washington DC 20059, USA
[2] African Institute for Mathematical Sciences, Limbe, Cameroon
mikel.ngueajio@bison.howard.edu, gloria.washington@howard.edu, danda.rawat@howard.edu, yolande@aims.edu.gh.

**Abstract.** With the growing rates of cyber-attacks and cyber espionage, the need for better and more powerful intrusion detection systems (IDS) is even more warranted nowadays. The basic task of an IDS is to act as the first line of defense, in detecting attacks on the internet. As intrusion tactics from intruders become more sophisticated and difficult to detect, researchers have started to apply novel Machine Learning (ML) techniques to effectively detect intruders and hence preserve internet users’ information and overall trust in the entire internet network security. Over the last decade, there has been an explosion of research on intrusion detection techniques based on ML and Deep Learning (DL) architectures on various cyber security-based datasets such as the DARPA, KDDCUP’99, NSL-KDD, CAIDA, CTU-13, UNSW-NB15. In this research, we review contemporary literature and provide a comprehensive survey of different types of intrusion detection technique that applies Support Vector Machines (SVMs) algorithms as a classifier. We focus only on studies that have been evaluated on the two most widely used datasets in cybersecurity namely: the KDDCUP’99 and the NSL-KDD datasets. We provide a summary of each method, identifying the role of the SVMs classifier, and all other algorithms involved in the studies. Furthermore, we present a critical review of each method, in tabular form, highlighting the performances measures, strengths, and limitations, of each of the methods surveyed.

## 1 Introduction

The internet has revolutionized communication and our entire way of life in such a way that we depend heavily on internet connectivity for almost everything we do. As a result, internet security has become one of the most important and relevant issues in today’s environment. Our increasing reliance on computer networks and related applications has also led to increasing and more sophisticated cyber intrusions, attacks, and espionage. To defend against these, lots of security techniques such as encryption and decryption mechanisms or Cryptography, firewalls, and IDS have been researched and applied over the years. Intrusion detection[1] is deemed the primordial line of defense against complex and dynamic intrusions as it builds upon existing models and patterns to detect, trace and stop intrusion[2]. The biggest problem with intrusion detection, however, is that to build effective models and behaviors, a massive amount of audit data needs to be collected. With the rapid increase in network traffic and exponential increase in network attacks and intrusions, better and more robust methods of data collection, analysis and classification are warranted to help expedite and facilitate patterns or signatures extraction from intruders and hackers.

To solve these problems, researchers have been looking for more sophisticated techniques for intrusion detection involving ML techniques. Machine Learning is a method of analyzing data that automates analytic model building. ML has gained popularity in cyber security especially in detecting network anomalies and has a proven track record of high detection rates [3]. There have been many implementations of ML algorithms in IDS including Naïve Bayes (NB) [4], Random Forests (RF) [5], Decision Trees (DT) [6], K Nearest Neighbor (KNN) [7], Logistics Regression (LR) [8]. In this research, we focus mainly on the implementation and applications of a special kind of ML algorithm called the SVMs [9].

SVMs are supervised ML models that can analyze data for both classification and regression problems. SVMs are particular in their capability of performing and generalizing quite well even on very small samples size training or learning data. Their robustness, versatility, and efficiency on datasets of all sizes make them a widely popular method, especially in intrusion detection [10]. Although there have been several existing literature surveying intrusion detections [11-13], very few have focused on using SVMs classifiers either individually or simultaneously with other algorithms. Additionally, quite a few literature studies use other cybersecurity datasets or a combination of multiple different cyber security-based datasets of comparative studies or reviews of IDS approaches involving multiple ML algorithms. Some popular datasets used in intrusion detection studies include;- the GTCS (Game Theory and Cybersecurity) dataset [14], DARPA-Lincoln [15, 16], UNSW-NB15 [8] datasets. An overview of various cybersecurity datasets, their characteristics, and ML usage and adoption for multiple cyber applications is presented in [17].

In this paper we present a comprehensive review of research in IDS, which applied SVMs classifiers singly, or combined with other methods, to detect or classify intrusions. We focus essentially on research evaluated and, tested on the KDDCUP'99 and/or the NSL-KDD datasets. The rest of the paper is structured as follows; In Sect. 2, we provide an overview of the SVMs algorithm, IDS, and a description of different kinds of IDS. We additionally discuss the significance of applying SVMs to intrusion detection. In Sect. 3, we give a brief overview of the datasets while Sect. 4 focuses on literature reviews of different research, techniques, and methods. In Sect. 5, we provide a tabular summary and critical review of the surveyed methods and Sect. 6 draws conclusions and potential future works.

## 2 Background

In this section, we provide more context, content, definitions, and explanation of key concepts used in this research. We also highlight the significance of SVMs in IDS, the different categories of IDS, and their pros and cons.

### 2.1 Support Vector Machines (SVMs)

SVMs are a set of related supervised learning methods used for both classification and regression problems. However, SVMs are mostly used in classification problems. SVMs do not make any assumptions about the data as their main objective is to find a hyperplane in an N-dimensional space that can distinctly classify the data points. Generally, there may exist many distinct possible hyperplanes to choose from, but the main goal is to find the plane with a maximum distance between data points of each class. Data points that fall closest to the hyperplanes, and hence help build the SVMs are called Support Vectors. In an SVMs classifier, it is easy to have a linear hyperplane separating different classes but in real-world scenarios, not all data can be linearly separable. So, in many cases, SVMs get associated with a function called the Kernel Trick, which does the job of transforming non-separable problems into separable problems. That is, the SVMs will use a kernel function that maps training vectors (with low dimensional space) to a higher dimensional space where a maximal separating hyperplane is designed, hence making them non-linear. The most common kernel functions used with SVMs [18] include: - the Gaussian Radial Basis Function (RBF), Polynomial, Sigmoid kernel, and Inverse multi-quadric kernel.

The Gaussian RBF kernel has shown to be the most effective kernel of all, mainly because SVMs often perform best on classification problems when associated with this kernel function [19]. Generally, in IDS the distribution of different types of attacks is imbalanced and the low-frequency attacks are usually too small compared to higher frequency attacks. For that reason, SVMs are quite effective and popular algorithms for intrusion detection particularly due to their good generalization capabilities even with smaller-sized datasets [19], their low classification latency [20], and their efficiency in classifying various classes [21].

### 2.2 Intrusion Detection Systems (IDS)

An intrusion is "any set of actions that attempts to compromise the Integrity, Confidentiality or Availability of resources" [16]. Intruders can get access to a system by exploiting weaknesses in the system either due to design and programming errors or simply through negligence by system supervisors. IDS are a set of software and hardware which attempts to spot computer attacks by examining various data audits and sorting them as either normal or malicious activities. The robustness of an IDS depends primarily on the strength of the intrusion detection technology used, and over the years, researchers have tried to find the best combination of accuracy and low modeling time. Unfortunately, traditional methods for intrusion detection are riddled with shortfalls such as low detection capabilities against unknown attacks, high false alarm rates, and insufficient analysis capabilities. Hence the drive toward AI for a more reliable solution to these problems [19].

Host-Based and Network-Based attacks are often classified as the two most common intrusion detection attacks. Host-based Intrusion Detection Systems (HIDS) work by monitoring, analyzing, and nefariously accessing the internals of an individual computer system as well as its network packets. On the other hand, Network-based attacks are comparatively more robust and are designed to make it complicated for intruders to gain access to the system, by purposely occupying and sabotaging network resources and services. Network-based Intrusion Detection Systems (NIDS) work by using network traffic data (i.e., Tcpdump) to scan through all traffic addressed to the machine being monitored [16].

The two main methods used to facilitate the detection of threats in IDS include: - the Signature-Based IDS and the Anomaly-Based or Knowledge-Based or Misuse IDS. Intrusion signatures are unique footprints or patterns that can be attributed to a specific cyber attacker. Signature-Based IDS are usually responsible for monitoring network systems in search of these intrusion signatures and then recording them so they can be investigated and analyzed by computer systems' administrators. This type of intrusion detection can be effective in processing high volumes of network traffic but are usually inept at detecting new or never-before-seen attacks as they can only identify patterns in known attacks [17]. Anomaly-based IDS techniques overcome these issues by first examining the behavior of the network to find any patterns, and then automatically creating a data-driven model to profile normal behaviors such that in case of an anomaly, the system would easily detect and recognize deviations from normalcy. The main advantage of Anomaly-based IDS is its ability to detect unknown and zero-day attacks [17, 22], since new legitimate behaviors can be falsely interpreted as attacks, hence resulting in high false positives. There exists a third type of IDS called the Stateful Protocol Analysis (SPA) with "Stateful" here meaning the IDS could know and trace protocol state, for instance, know how to pair requests with replies. SPA processes are quite similar to Anomaly-based IDS, but for the fact that Anomaly-based IDS adopt a preloaded network or host-specific profiles, whereas SPA depends on vendor-developed generic profiles to specific protocols [22]. Table 1 shows the pros and cons of Signature-based IDS, Anomaly-based IDS, and SPA.

**Table 1**. Different Types of IDS, their Pros and Cons.

| Types of IDS | Pros | Con |
|---|---|---|
| Signature-based IDS | -Simplest and most effective for<br>detecting known attacks.<br>-Detail Contextual Analysis. | -Hard to keep track of patterns.<br>-certain tasks may be time consuming<br>-Ineffective at detecting<br>unknown attacks, variants of known, and evasion attacks. |
| Anomaly-based IDS | -Effective at detecting<br>new or unforeseen attacks.<br>- Less dependent on OS<br>-Can easily detect privilege Abuse | -Weak profiles accuracy due to constantly changing events<br>-Unavailable during behavior profiles design.<br>-Difficult to trigger an alert at the right time. |
| Stateful Protocol Analysis (SPA) | -Can easily distinguish unexpected sequences of commands<br>-Can recognize and trace the protocol states | -Can be hard to protocol state tracing and examination.<br>-Ineffective at inspecting attacks that resemble benign protocol behaviors. |

## 3 Datasets Descriptions and Overview.

This section provides a brief overview of the KDDCUP'99 and NSL-KDD datasets, highlighting some of their features and most important characteristics.

### 3.1 The KDDCUP'99 Dataset

The KDDCUP'99 data set is a widely used dataset for building IDS [23], and for evaluating anomaly detection methods were in addition to the normal intrusion category (networks that do not contain any intrusions), attacks are categorized into either one of four classes [17]: - Denial Of Service (DoS): Shutting down a network by flooding it with information and requests; Remote-to-Local (R2L): Unauthorized access from a remote machine; User-to-Remote (U2R): Intruder attempts to gain access to a normal user account; Probe: Surveillance intrusions on networks.

The KDDCUP'99 is a benchmark dataset in cybersecurity, created in 1999 by the Defense Advanced Research Project Agency (DARPA). It was used for the Third International Knowledge Discovery and Data Mining Tools competition whose main goal was the creation of a network intrusion model capable of classifying intrusions as either good, bad, or normal. The dataset comprises 4,898,430 samples each containing 41 features belonging to one of 3 groups namely: - Basic features, Content features, or Traffic features, further divided into Time-based or Host-based traffic features. In this dataset, a data point is either normal or an attack, and each attack can be labeled as either a DoS, U2R, R2L or a Probing Attack. Despite its wide application, this dataset has some inherent flaws which may contribute to its low performance, long training time, and overall poor evaluations of intrusion detection approaches. The dataset is heavily skewed, and about 78% and 75% of records in the train and test set respectively are duplicated records [24]. Table 2 presents a summary statistic of redundant records and classes imbalance associated with the KDDCUP'99 dataset.

**Table 2.** Summary statistics of redundant records in the KDDCUP'99 Dataset.

| Sets | Traffics | Original record | Distinct record |
|---|---|---|---|
| Training set | Attacks | 3 925 650 | 262 178 |
| | Normal | 972 781 | 812 814 |
| | Total | 4 898 431 | 1 074 992 |
| | Attacks | 250 436 | 29 378 |
| Testing Set | Normal | 60 591 | 47 911 |
| | Total | 311 027 | 77 289 |

### 3.2 The NSL-KDD Dataset

The NSL-KDD dataset was proposed in 2009 as a refined version of the KDDCUP'99 dataset and advent to solve some of its inherent problems. Primarily, the NSL-KDD dataset is comparatively smaller in size, mainly due to the removal of all duplicate records in its training and test sets. Furthermore, its simplification led to it being more manageable and thus, can easily be used in its totality. Consequently, there is little chance of models' bias due to records' skewness. Also, to ensure that the categorization percentage for different ML models fluctuates widely, the number of sampled records from each difficulty level was chosen such that they were inversely comparable to the ratio of records in the original KDDCUP'99 dataset.

The NSL-KDD dataset has most of the inherent properties of its parent dataset. For instance, every data point is labeled as either normal or an attack, with each attack either being a DoS, U2R, R2L, or a Probe. It also comprises 41 features (3 nominal, 4 binary, and 34 continuous) categorized as either Basic features, Content features, or Traffic features, with 23 traffic classes in the training set and 30 in the testing set.

## 4 Literature Review

In this section, we present different methods proposed over the years, employing SVMs classifiers to build better and more powerful IDS, and implemented or tested either on the KDDCUP'99 dataset, NSL-KDD dataset, or both datasets.

### 4.1 IDS using SVMs on the KDDCUP'99 Dataset

Among the variety of approaches, SVMs are one of the most used ML algorithms in intrusion detection. They are often used independently or in combination with other algorithms or methods but are mostly employed in classifying attacks. In IDS, features are sometimes redundant or less influential in separating data points into correct classes however, the performance of a model may sometimes depend on selecting the best parameters for the task. As a result, methods that provide optimal feature selection and model optimization are often used together with the SVMs model as they contribute to either enhancing training time, reducing false positives and false negatives, and/or boosting the overall SVMs model performances.

A novel method for IDS using a combination of 3 algorithms: - Kernel Principal Component Analysis (KPCA), SVMs, and Genetic Algorithm (GA) is proposed by [19]. In this study, the KPCA is used as a preprocessor for the SVMs, to help reduce the features' dimension space by extracting the most effective subset of all attributes and discarding the rest. The research experiment involves using four SVM classifiers to identify the Normal, DoS, R2L, U2R, and Probing attack states, with each classifier associated with an N-RBF kernel function. The N-RBF function is efficient at reducing training time and was also developed to help reduce the noise caused by feature differences. The selected features are then fed into the GA algorithms, whose role is to find the most optimal parameters for the SVMs classifier. Experimental results illustrate better generalization, high predictive accuracy with a max detection rate of 96.38% and false alarm rate of 0.96, and a faster convergence time. However, this method is quite complex for real-life applications. A similar method is proposed by [25], but for the addition of an Improved Chaotic Particle Swamp Optimizer (ICPSO). Here, the KPCA is used to reduce the feature space while the ICPSO is used to select suitable parameters for the SVMs classifier, which introduces chaos optimization, premature judgment, and processing mechanism. This method produces an optimized and more efficient SVMs classifier with shortened training time, but the method shows a poor detection rate forecasting attacks such as R2L and U2R.

The research proposed by [26], involves a 6-part model architecture (data selection, feature transformation, feature subset selection, classification, training and testing, and, results) and is implemented using 3 algorithms: - PCA, GA, and SVMs. Instead of employing the conventional approaches of utilizing PCA for selecting features with the highest eigenvalues, even at the risk of these features not being optimally sensitive to the classifiers, this method utilizes the GA to search through the principal component feature space and extract highly sensitive subgroup of features with high discriminatory power. The PCA is then used for feature transformation, and to reduce data redundancy while the SVMs is used to classify attacks. The combination of PCA and GA for feature selection provides for a simpler (minimum number of features:10), more efficient, and more accurate SVMs classifier (Maximum detection rate of 99.96%) compared to other approaches. The downfall is the method's complexity and its performance, which can further be improved by raising the recognition rates and diminishing false positives [27].

There have been quite a few studies in IDS using GA and SVMs but most fail to exploit the potential benefits of utilizing both models together. An enhanced anomaly detection technique based on simultaneously using SVMs with GA is proposed in [28]. The results illustrated a high accuracy of 98.33% with the hybrid model compared to 94.80% using SVMs alone, 99.50% true positive rate, and 98.22% true negative rates. The paper proves that enhancing SVMs with GA results in reduced false alarms and mean square errors.

Despite their popularity, SVMs can be computationally demanding so, dimension reduction and other optimization techniques are often introduced during training, to help extract the best features for the model, and possibly achieve maximum accuracy.

A network intrusion detection method using SVMs with RBF kernel is introduced in [29]. The research presents ways to choose, and optimize training features for SVM models, and test this approach with an experiment that involves detecting intrusions on an open network IDS called SNORT in an online setting, using SVMs with RBF kernel. K fold Cross-Validation is implemented for model optimization, and to help select the best attribute for the RBF kernel parameters - gamma and the penalty coefficient cost C. Experimental results show a significant decrease in false positive and negative rates and faster detection speed. The best-performing model achieves a 96.6543% accuracy. However, this method may be computationally taxing. The investigation of [30]'s utilizes Linear SVM to categorize attacks as either normal or abnormal. Initially, the data is preprocessed by encoding the categorical features, discarding missing values, data splitting into training and test sets, and finally, feature scaling using the standard scaler method. The SVM model is then trained, and its performance is evaluated using precision, recall, F1_score, and a confusion matrix. In the end, considering all 41 features from the original dataset, the Linear SVM model achieves an accuracy, precision, recall, and F1_score of 96.6%, 98.0%, 95.3%, and 96.7% respectively, better than earlier research that employed this same technique. A novel method proposed by [31] involves the creation of an ensemble-based classifier by training and testing 12 "Experts" – 6 SVMs experts and 6 KNN experts and eventually combining them into a single model. Three ways are used to combine these experts' opinions and involve either generating weights using either PSO or from a Meta-optimized PSO, or by using a Weighted Majority Voting (WMV) approach. Initially, the data is processed by converting all features and attack classes (DoS, Normal, U2R, R2L, and Probe) into numeric values. Then, 5 binary classifiers were trained on the KNN experts and 5 others on the SVMs experts, one for each class. The weights for each of the classifier is generated using different variants of the Principal Swamp Optimization (PSO) technique, and the WMV is used to combine the opinions of the base experts into an ensemble model. The method with PSO produces a 0.756% performance boost compared to the accuracy of the best performing base expert. However, training all the experts may be computationally expensive. Likewise, [32] proposes a combination of PSO with SVMs. They first separate the training dataset into normal and attack classes, which are then fed into the feature selection algorithms- the PSO to select the best and most important features for the task. The selected features are then used to build and train the SVMs model and the results from this research show optimal feature selection through high detection rates of 99.8438% compared to 82.6387% for the approach not employing feature selection.

An intelligent approach for anomaly intrusion detection using SVMs, Decision Trees (DT), and the Simulated Annealing (SA) is presented by [33]. In this study, the SVMs and SA are used together to find the best candidate features for intrusion detection attacks while the combination of the DT and the SA is used to produce and obtain decision rules for new attacks and to boost overall model performance. The SA method is also used automatically to set the best parameters for the DT and the SVMs. The proposed method also utilizes a ten-fold Cross-Validation (CV) to help create a robust model. The result shows 99.96% accuracy with minimal selected features compared to other State-Of-The-Art (SOTA) approaches. The author in [34] also proposes a method for intrusion detection using a Decision-Tree-based SVM algorithm. Here, the DT model consists of a series of 5 binary class SVMs, constructed by pairing single attacks from different attack classes (DoS, U2R, R2L, Probe) with normal attacks, and the final structure of the DT model is influenced by the remoteness of binary class patterns and by the size of each class patterns. The DT is constructed from root node to leaf node and is used to classify intrusion, by iteratively comparing intrusion that needs to be detected with the root node of the tree. An SVM with RBF kernel (parameters set $c= 1$, $\varepsilon= 0.1$, and $\gamma= 0.5$) is used and the results show a detection accuracy of 96% with low false positive and false negative rates.

An efficient and reliable classifier build by combining SVMs, Ant Colony Algorithm, and K-means clustering to detect normal and abnormal behaviors in a system is presented in [35]. The proposed method applies K-Means clustering to reduce the size of the dataset from 145 586 samples to 116 266 samples and the Ant Colony Algorithm to help choose the proper training subset by randomly sampling all subsets and selecting the most representative of the whole network. The sampled subset is then trained and tested using the SVMs with a ten-fold CV. Overall, the model achieves a high accuracy of 98.6249% and an average Matthews Correlation Coefficient (MCC) of 0.861161 for intrusion detection. This method however is not adaptive to multi-class problems, especially in unbalanced circumstances, and may also be resource-demanding.

Some hybrid techniques involving Neural Networks (NN) and SVMs have been proposed and found to be quite effective in detecting attacks. For instance, [36] explore the feasibility of applying a hybrid combination of Artificial Neural Network (ANN, feed-forward NN trained with Back Propagation (BP) algorithms) with SVMs to predict attacks based on frequency-based encoding techniques. The main objective of this research is to investigate which of the two methods is capable of generalizing and performing well even with limited training data. For the experiment, the dataset is first converted to a specific vector space representation, then the training sample is partitioned into the normal and attack classes, which are subsequently fed into both algorithms. The ANN and SVMs predictive models are later evaluated on the tests sets and a ROC curve is built to visualize the results for various detection thresholds. At the end of the experiment, it is observed that the SVMs model outperforms the ANN model with an attack detection rate close to 100% for each but false-positive rates of 10% and 40.7% for the SVMs and ANN models respectively. Similarly, [37]'s approach to intrusion detection, involves the use of SVMs and NN (multi-layer, feed-forward networks) on normal user activity and attacks patterns, so, the models can learn to detect and easily recognize between anomalies and Misuse IDS. The idea is to train a NN-based, and SVM-based models to detect and learn these patterns so that any deviations from normal behaviors are easily identified and flagged as attacks. A comparative study of these two algorithms shows that the SVMs-based model outperforms the NN-based algorithm with about a 99% accuracy on the test set. In addition to that, there is a significant difference in training time between them, 17.77 seconds for the SVMs-based model versus 18mins for the NN-based model. However, this method could be CPU intensive and require a large amount of data to perform best, as a result of modeling with deep learning architectures.

### 4.2 IDS using SVMs on the NSL-KDD Dataset

In this subsection, we describe some relevant research on intrusion detection using SVMs, applying the NSL-KDD dataset as the main training and testing use case.

A recent study [38] employs SVMs with RBF kernel, RF, DT, and Multi-Layered Perceptron (MLP) to classify intrusions as either normal or malicious. First, data preprocessing involves deleting unnecessary features, and handling missing values; then, data normalization and standardization are achieved, Then, the data is split into training and validation sets and applied to each model separately. The performance of each model was evaluated on three different subsets of features from the NSL-KDD dataset. Research findings show that the SVMs classifier achieved an average accuracy of 98% on all three subsets, outperforming the MLP and DT classifiers. However, to reduce computation time, only a sample of the original dataset was used in the experiment. A novel technique designed to produce faster training times is proposed by [39]. In this approach, a series of data preprocessing techniques are applied to the dataset, to prepare it for modeling. The dataset transformation phase helps convert features to numerical values; then there is dataset normalization using the min-max method, to help boost SVM model performance, and lastly, dataset discretization for continuous feature selection of intrusion detection parameters and consistency between values. The experiment uses multiple SVMs kernel functions namely: - RBF, polynomial, and sigmoid kernels, with a 10-fold CV, as well as a CV with reevaluation method. The authors report an increase in accuracy from 94.1857% with CV to 98.5749% with CV and re-evaluation, with both models requiring the same amount of time (77.01 seconds) to build and train the SVMs with RBF kernel. Thus, validating the observation that proper data preprocessing and feature selection are essential for increasing the attack detection rates for SVMs kernel models. In the same vein, [40] present four types of SVMs model kernel-based intrusion detection techniques involving the Linear, Quadratic, Fine Gaussian, and Medium Gaussian SVMs. In this method, a five-fold CV scheme is used for validation and the SVMs' role is mainly to classify the intrusion as either normal or malicious attacks. The technique involving the Linear, Quadratic, Fine Gaussian, and medium Gaussian SVMs results in 96.1%, 98.6%, 98.7%, 98.5%, accuracies, respectively, with errors of 3.9%, 1.4%, 1.3%, 1.5% respectively, and 74s, 53s, 100s, 50s model building times respectively. The study concludes that the Fine Gaussian SVM performs the best, however, the method could use an SVMs optimization technique to enhance its overall performance.

A comparative study of eight ML algorithms: - SVM with RBF kernel, KNN, LR, NB, MLP, RF, Extra Tree Classifier (ETC), and DT for classifying intrusion as either normal or malicious, is proposed by [41]. The training and testing phases employ different random selections of the five attack classes namely, Normal, DoS, R2L, U2R, and Probe extracted from the original set. Only significantly important attributes are selected for this task with hopes of boosting performance, reducing modeling overheads and training time, and are categorized into four groups. After data preprocessing, the classifiers were trained and tested on 111 386 and 37 129 instances respectively, and the performance of each model is evaluated on accuracy, precision, recall, and F1_score, for each of the classes. SVMs do not perform among the top classifiers in this study, with an average accuracy of 88% but show promising results for DOS attacks detection. On the other hand, [42] applies a Recursive Feature Elimination method (RFE) to help select relevant features and conducts their experiment using Random Forests (RF) and SVMs. In addition to that, a comparative study of both methods before and after feature selection is presented. In the data preprocessing phase, all categorical features from the training and test set are transformed to numerical values using one-Hot-Encoding (OHE), all missing categories are discarded, and each algorithm is evaluated using Confusion Matrices. Only the best 13 features are selected using the RFE method and used by both classifiers and the SVMs outperform RF after feature selection for most attack classes. The approach proposed by [43] consists of merging feature selection with multi-class classification using SVMs. The idea is to significantly reduce the number of features required in the modeling process while maintaining the model's performance. This approach applies SVMs on multiple input feature subsets of the training set, and the result show 91% classification accuracy using only 3 features and 99% accuracy with 36 and 41 features. Similarly, [44]'s methodology employs SVMs for feature ranking and classification analysis. However, optimal parameters for the SVMs model are obtained through Grid Searching, and an RBF kernel is used with the SVMs for intrusion classification. In this approach, the Information Gain (IG) Feature Ranking method is used to obtain a reduced set of features for the SVM, and in the end, 41 features are used for analysis. The result of this analysis shows a much improved SVMs performance compared to the original benchmark results for this dataset. Another method involving IG Ratio, alongside a combination of K-means and SVMs is proposed by [45]. In this approach, the feature selection and ranking are done using the IG ratio while the K-means classifier is used to compute the detection rate for each subset of features selected. These features are then used to train the SVMs classifier and the results for the reduced dataset show shorter computational time, increased detection accuracy, and performance. The investigation presented by [46] also involves IG for Feature selection, with an SVMs classifier optimized with Swarm Intelligence. The SVMs optimized with an Artificial Bee Colony (ABC) method are used for intrusion classification and the approach is implemented on WEKA. The result is compared with the standard SVMs (with default options) and overall, the method shows a lower false alarm (0.03) and higher detection rate (98.53) compared to regular SVMs. However, only two swarm intelligence algorithms are implemented.

An efficient intrusion detection task for Wireless Sensor Networks (WSN) is proposed by [47]. The author utilized several data transformation techniques such as the Principle Component Analysis (PCA), Linear Discriminant Analysis (LDA), and Local Binary Pattern (LBP) to transform the features into principle feature spaces. PSO is then applied to select the most optimal subset of features based on their sensitivity and high discriminatory powers. Finally, the SVMs and MLP are employed for intrusion classification. Research outcomes prove that this method is capable of providing optimal intrusion detection mechanisms with minimal features, and maximum detection rates. Another Intrusion detection technique for WSNs proposed by [48] involves using a Modified Binary Grey Wolf optimizer with SVM (GWO-SVM). The model comprises training the SVMs with a combination of 3, 5, and 7 wolves, to find the optimal number of wolves necessary to obtain the best model performance while constantly adjusting the fitness function which validates the feature subset considered for the task. Research experimentation comprises categorical features encoding, feature normalization, and feature selection using the modified GWO. Then a restrained dataset is trained on the SVM model whose performance is evaluated in terms of the number of features used, model accuracy, training time, false alarm, and detection rate. The authors present a comparative study of other feature selection methods such as PSO and results show that the GWO-SVM method with 7 wolves achieves a 96%, 3%, and 96%, accuracy, false alarm rate, and detection rates respectively, with only 17% of the features selected.

The negative effects of dataset shift in intrusion detection are addressed in [49] and a technique to solve this problem involving a modifier SVMs and Neural Network (NN) Back-Propagation scheme using Covariate shift is presented. In the experiment, the SVMs model is modified using the Kernel Mean Matching (KMM) and Unconstrained Least-Squares Importance Fitting (ULSIF) technique while the Back Propagation (BP) model is modified using the LaGrange multipliers and a specified loss function, which is then used to update the weights of the Neural Network. A thousand samples of data are used for training and validation respectively, while 22 544 data points are used for testing. The result shows higher accuracy for modified techniques (82.50% and 82.60% for SVMs and BP respectively) as compared to the nine best original techniques studied.

### 4.3 IDS using SVMs on both the KDDCUP'99 and the NSL-KDD Dataset

Over time, much research has utilized AI and ML algorithms to classify, detect and identify intrusion detection using either the NSL-KDD dataset or the KDDCUP'99 dataset. Despite the huge shift from KDDCUP'99 to NSL-KDD over the years, SVMs are still used in comparative analysis involving both datasets to determine which performs best in intrusion detection studies. In this section, we describe some research aimed at comparing the quality and efficiency of the SVMs algorithm on both the KDDCUP'99 and the NSL-KDD datasets.

The method proposed by [46] first converts all the categorical features in both datasets using OHE and normalizes values between 0 and 1 using the Euclidean normalization. In their comparative study, the authors use the ANN, SVMs, Naïve Bayes, and RF, on both datasets. At the end of their experimentation, the RF model achieved the best precision of 99.9% for the KDDCUP'99 and 99.0% for the NSL-KDD dataset, while SVMs achieves the best F1_score of 95.8% for the KDDCUP'99 and 84.8% for the NSL-KDD dataset. In this other study by [50], a novel method called a few-shot DL approach is presented. The result shows that the proposed method achieves better performance than the SOTA results for these two datasets. The idea presented in this research is to train a Convolution Neural Network (CNN) on intrusion detection, then extract outputs from different layers of CNN while implementing Linear SVMs with a 1-Nearest neighbor classifier for intrusion detection. This method obtained a state-of-the-art performance on the KDDCUP'99 and NSL-KDD datasets with over 94% accuracies on both datasets.

## 5 Critical Analyses of Different Approaches Surveyed.

This section presents a summary review of the different approaches surveyed in tabular form. In Table 3 below, we provide a critical review of the surveyed research involving SVMs for IDS by summarizing the proposed methodology, emphasizing the dataset used, (abbreviated DS, while the KDDCUP'99 and NSL-KDD datasets are KDD and NSL respectively), the role of the different models involved, and some strengths and/or weaknesses of the approaches.

**Table 3.** Critical Review of IDS using SVMs on KDDCUP'99 and/or the NSL-KDD datasets

| Ref / DS | Proposed Method | Approach or solution | Role of SVMs | Strength | Weakness |
|---|---|---|---|---|---|
| [19] KDD | KPCA, SVMs, and Genetic Algorithm | -KPCA to reduce the dimension of the feature -GA for optimal parameters selection for the SVMs. | -Intrusion detection & classification | -better performance with a 96.38% max detection rate and 0.96 false alarm rate. Faster training time and testing time. | Complex for real-time application. |
| [25] KDD | KPCA, SVMs, and ICPSO | -KPCA for feature dimension reduction -ICPSO for best parameters selection for the SVM classifier. | -Intrusion detection & classification | -Improved SVMs model with faster computational time, -High predictive accuracy. | -Poor detection rate for attacks like R2L and U2R. |
| [26] KDD | PCA, GA, and SVMs. | -GA searches the genetic principal component features with optimal sensitivity -PCA for data transformation redundancy reduction. | -Intrusion detection & classification | -A simpler, more efficient, and more accurate SVMs classifier. -Maximum detection rate of 99.96% compared to state-of-the-art | -Complex Performance issues can be improved by raising the recognition rates |
| [28] KDD | hybrid SVMs via GA. | GA is used as optimizer of intrusion recognition system | -Intrusion detection & classification | -High accuracy of 98.33%. - True positive rate 99.50% and 98.22% True negative rates | - |
| [29] KDD | SVMs with RBF Kernel | -K fold CV to help select the best attributes for model optimization | -Intrusion detection, classification, and Feature selection | -Reduced false negative and false positive -Faster detection -Best accuracy of 96.65% | computationally taxing and time-consuming |
| [30] KDD | Linear SVMs | Feature scaling with a Standard Scaler metric | Intrusion classification | Accuracy, recall, precision, and F1-score of 96.6%, 98%, 96.7%, and 95.3% respectively | - |
| [31] KDD | SVMs, KNN, and Particle Swamp optimization (PSO) | -Combines 6 SVMs and 6 KNN, and trains them on each attack class, and the weights of each classifier are used to combine the opinions of base experts. | -Improves accuracy of intrusion detection -Attack classification | -Best accuracies greater than 88%. | Training all the experts' may be computationally demanding |
| [32] KDD | PSO + SVMs | -Creates an attack and normal class. -PSO for feature selection, which is then used by the classifier | -Attacks classification into normal or abnormal | Fewer features are used in training models hence higher detection rates of 99.8% | - |
| [33] KDD | DT + SVMs + Simulated Annealing (SA) | -SVMs and SA for best feature selection. -DT and SA are used as decision rules for new attacks. -SA sets the best parameters for DT and SVMs. | -Feature selection -Attack classification -Intrusion detection | -Minimal features to train models -High accuracy of 99.96% -Better results compared to winning results for the KDDCUP'99 competition. | - |
| [34] KDD | DT and SVMs with RBF Kernel | -Decision tree comprises 5 bi-classes. SVMs constructed by pairing an attack class with a normal class. | -Intrusion classification -Better average performance | -Lower false positive rates -Lower false negative rates High detection accuracy of 96% | - |
| [35] KDD | K- Means + Ant Colony + SVMs with 10 folds CV | -K-Means clustering for feature reduction -Ant colony for best training subset selection | -Attack classification -Intrusion detection. | -High accuracy 98.6% -Reduced features used in modeling hence faster. - MCC of 0.86 | -Not adaptive to multi-class imbalanced problems |
| [36] KDD | Artificial Neural Network + SVMs | -Based on a high frequency-based encoding method -data fed into both models separately and result compared. | -Attack classification -Intrusion detection | -High Attacks detection rate. -SVMs Lower False positive rate 10%. ANN has a false positive rate of 40.7% | CPU intensive due to the use of deep learning techniques |
| [37] KDD | Neural networks (NN) + SVMs | -Comparative study to see which detects more intrusions. -Use both models on normal user activities, flag any deviation from normal | -Normal pattern recognition and anomaly detection | -High accuracy of 99% for SVMs. -Reduced training time for SVMs - 17.7s compared to 18 Mins for the NN model | May be CPU intensive due to NN architecture and techniques used |
| [38] NSL | SVMs with RBF kernel, RF, DT, and MLP | Comparative study of multiple ML algorithms for Intrusion classification | Intrusion classification | SVMs with RBF kernel achieve 88% accuracy, outperforming MLP and DT | Only a sample of the original dataset is used |
| [39] | SVMs with different Kernel functions | Extensive Data preprocessing to prepare for modeling and use of different SVMs | -Attack classification -Intrusion detection. | - SVMs +CV with reevaluation, High accuracy of 98.6% -High attack detection rate | |
| NSL | + 10 fold CV | kernels namely, Polynomial, RBF, Sigmoid, 10-fold CV, and CV with reevaluation. | | -Shorter computation time and less model building time. | - |
| [40] NSL | Linear, Quadratic, Fine, and Medium Gaussian SVMs | A 5-fold CV scheme is used for validation and SVMs classify the intrusion as either normal or malicious attacks | Intrusion classification | 96.1, 98.6, 98.7 and 98.5% detection accuracies, respectively, and 3.9, 1.4% 1.3% and 1.5%, errors respectively | Requires SVMs optimization technique to perform best |
| [41] NSL | SVMs, RBF, LR, KNN, RF, NB, LP, ETC, and DT | -Comparative study of multiple ML algorithms for Intrusion classification -Optimal feature selection method | Intrusion classification | SVMs achieve average Accuracy of 88% with promising results for DoS attack detection | Modeled on a small subset of the original dataset |
| [42] NSL | RFE, SVMs. and RF | -RFE for feature selection (FS) -Comparative study of SVM and RF performance before and after FS | Intrusion classification | SVMs outperform RF after feature selection for most attack class | A confusion matrix is the only metric evaluator used |
| [43] NSL | SVMs | involves merging feature selection and multi-class classification fusing SVMs. Thus, requires fewer features for modeling | -Feature selection -Feature size-reduction -Intrusion classification | -High accuracy measure of 91% using only 3 features. -96% accuracy with 36 and 41 features | - |
| [44] NSL | SVMs with RBF and Grid Searching, IG feature ranking | -Reduced features obtained from IG feature ranking. -Grid searching to obtain the best parameters. | -Attack classification , intrusion detection, -Faster processing time | -F_measure results from this method outperform State-of-the-art results in [24] | - |
| [45] NSL | Information Gain (IG) Ratio + K-means + SVMs | -IG Ratio for feature selection and ranking. -K-Means help to compute the detection rate for each feature subset, which are | -Attack classification and intrusion detection -Best performance or detection accuracy | -Reduced computation time -Increased detection accuracy and performance. -Best accuracy of 99.37% and 99.32% with 30 and 23 | - |
| | | then used by the SVMs classifier. | | features respectively. | |
| [46] NSL | IG feather selection and SVMs classifier | -IG is used for feature selection | -SVMs optimized with swarm intelligence for intrusion classification | -lower false alarm 0.03 -higher detection rate of 98.53% toward regular SVMs | -Limited experimentation -few experiment implemented |
| [47] NSL | CA+ PSO + SVMs + MLP | -CA for feature transformation. -PSO for optimal feature selection -MLP used for modeling | -Optimal intrusion detection -Attack classification | -Optimal intrusion detection mechanism -Minimal Feature -Maximal detection rates | - |
| [48] NSL | Modified Binary GWO with SVM | GWO with 7 wolves applied for optimal feature selection and reduced execution time | -intrusion detection | The method achieves 96%, 3%, and 96% accuracy, false alarm, and detection rates respectively, with only 17% of features selected. | Complex and resource demanding. |
| [49] NSL | Modified SVMs and Backpropagation (BP) | -SVMs are modified using KMM and ULSIF technique -BP is modified using the LaGrange multipliers and loss function | Solve the dataset shift problem in building IDS | -Accuracy: 82.50 % and 82.60% for SVMs and BP respectively, -Avoids the curse of dimensionality -easily applicable in practice | - |
| [50] Both | SVMs, Naïve Bayes, Random Forest, and Artificial Neural Networks | -Converts all categorical features to numerical, using OHE. -Feature normalization using Euclidean normalization. Compare four ML algorithms on both datasets. | -Intrusion detection and classification | -Random Forest achieved a precision of 99.9%, 99.0% on the KDD and NSL respectively -SVMs have the best f-measure with 95.8% for KDD and 84.8% for the NSL dataset. | May be CPU intensive due to deep learning architecture and techniques used |
| [51] | CNN and the Few shots detection approach | -CNN to extract output features from its layers. -Feed these features into a | -Method for improving intrusion detection performance | High accuracy of 94% on both datasets | Possibly CPU intensive due to deep learning |
| Both | (Linear SVMs + 1-Nearest Neighbor classifier) | linear SVMs with a one-nearest neighbor classifier, for a few-shot detection. | | | architecture and techniques used |

## 6 Conclusions and Future works

Rapid advances on the internet coupled with modern communication and social media have led to an explosion of network data and consequently, record numbers of novel cyber security threats. These recent trends have galvanized more researchers into inventing and implementing more sophisticated approaches, employing ML algorithms such as SVMs to help counter these menaces to our online activities. SVMs are considered one of the pioneer ML algorithms for intrusion detection mainly due to their outstanding generalization power irrespective of dataset size, and their ability to overcome the curse of dimensionality. Of the plethora of cybersecurity datasets out there, we focus primarily on researches that have adopted SVMs and implemented their approach on the two most popular intrusion detection datasets namely: - the KDDCUP'99 and the NSL-KDD datasets

In this review paper, we have described IDS and provided background information on the different kinds of IDS and SVMs. We also discussed the importance of using SVMs in intrusion detection and provided a detailed description of the KDDCUP'99 and the NSL-KDD datasets. We have reviewed research using SVMs for intrusion detections, their approaches, and methodology. Additionally, we have presented a tabulated summary and critical analysis of each of these approaches, evaluated on either the KDDCUP'99, the NSL-KDD datasets, or both datasets, highlighting the roles of the different algorithms used, as well as the strengths, and weaknesses of the approach.

In future works, we will focus on expanding the current study to include systematic analysis and review of IDS implemented on other well-known cybersecurity datasets like the DARPA or the UNSW-NB15 using State-Of-The-Art Machine and Deep Learning models such as the Artificial Neural networks, K-Mean Clustering, SVMs.